\documentclass[conference]{IEEEtran}

\usepackage{cite}
\usepackage{amsmath,amssymb}
\usepackage{graphicx}
\usepackage{booktabs}
\usepackage{multirow}
\usepackage{xcolor}
\usepackage{url}
\usepackage{placeins}

\renewcommand{\baselinestretch}{1.004}

\newif\ifanonsubmission
\anonsubmissionfalse

\title{Rethinking Feature Conditioning for Robust Forged Media Detection in Edge AI Sensing Systems}

\ifanonsubmission
\author{\IEEEauthorblockN{Anonymous IWCMC Submission}}
\else
\author{
\IEEEauthorblockN{
Izaldein Al-Zyoud\IEEEauthorrefmark{1}, Member, IEEE, and
Abdulmotaleb El Saddik\IEEEauthorrefmark{1}, Fellow, IEEE
}
\IEEEauthorblockA{
\IEEEauthorrefmark{1}MCRLab, School of Electrical Engineering and Computer Science, University of Ottawa, Ottawa, ON, Canada \\
Corresponding author: Izaldein Al-Zyoud (e-mail: izzy.alzyoud@uottawa.ca).
}
}
\fi

\begin{document}
\maketitle

\begin{abstract}
Generalization under manipulation and dataset shift remains a core challenge in forged media detection for AI\mbox{-}driven edge sensing systems. Frozen vision foundation models with linear probes are strong baselines, but most pipelines use default backbone outputs without testing conditioning at the frozen feature interface. We present the first controlled probing study on DINOv3 ConvNeXt and show that, without task-specific fine-tuning, linear probing alone yields competitive forged-media detection performance, indicating that ViT-7B self-supervised distillation transfers to security-critical vision workloads at edge\mbox{-}compatible inference cost. Backbone, head, data, and optimization are fixed while conditioning is varied; LN-Affine, the default ConvNeXt head output, is the natural baseline. On FaceForensics++ c23, five conditioning variants are evaluated under in\mbox{-}distribution testing, leave-one-manipulation-out (LOMO), and cross-dataset transfer to Celeb-DF v2 and DeepFakeDetection. In ConvNeXt-Tiny, conditioning alone changes LOMO mean AUC by 6.1 points and reverses ID-vs-OOD ranking: LN-Affine is strongest on external datasets, while LayerNorm is strongest in\mbox{-}distribution. In ConvNeXt-Base replication, the OOD winner becomes protocol-dependent, and ID-optimal selection still fails as a robust deployment rule. Results show that feature conditioning is a first-order design variable and should be selected with robustness-oriented validation, not ID accuracy alone.

\end{abstract}


\section{Introduction}
The integrity of vision perception modules in AI-driven sensing systems depends on the authenticity of their visual inputs. In industrial inspection, smart-city surveillance, and intelligent connected vehicle (ICV) systems, vision-driven decisions are increasingly automated with limited human oversight. Forged synthetic media is therefore a direct operational threat: manipulated content can corrupt downstream decisions, weaken biometric verification, and reduce trust in AI-enabled security infrastructure \cite{amerini2025deepfake}.

Face deepfake video detection is a concrete, well-benchmarked instantiation of this broader forged-media threat. Yet generalization under distribution shift remains a core failure mode: methods that perform well in-distribution (ID) can degrade under unseen manipulations or external datasets \cite{rossler2019faceforensics,yan2023deepfakebench}. Foundation-model pipelines with frozen backbones and linear probes are attractive for this setting because they are simple, scalable, and practical for edge deployment \cite{ojha2023ucf,oquab2024dinov2,simeoni2025dinov3}, but they typically consume default backbone outputs without testing how post-pool conditioning affects robustness.
This setting is operationally important because DINOv3 ConvNeXt can be deployed without task-specific fine-tuning: a linear probe over frozen features is already competitive, suggesting that ViT-7B-distilled representations transfer effectively to security-critical edge vision tasks.

\begin{figure*}[t]
\centering
\includegraphics[width=\textwidth]{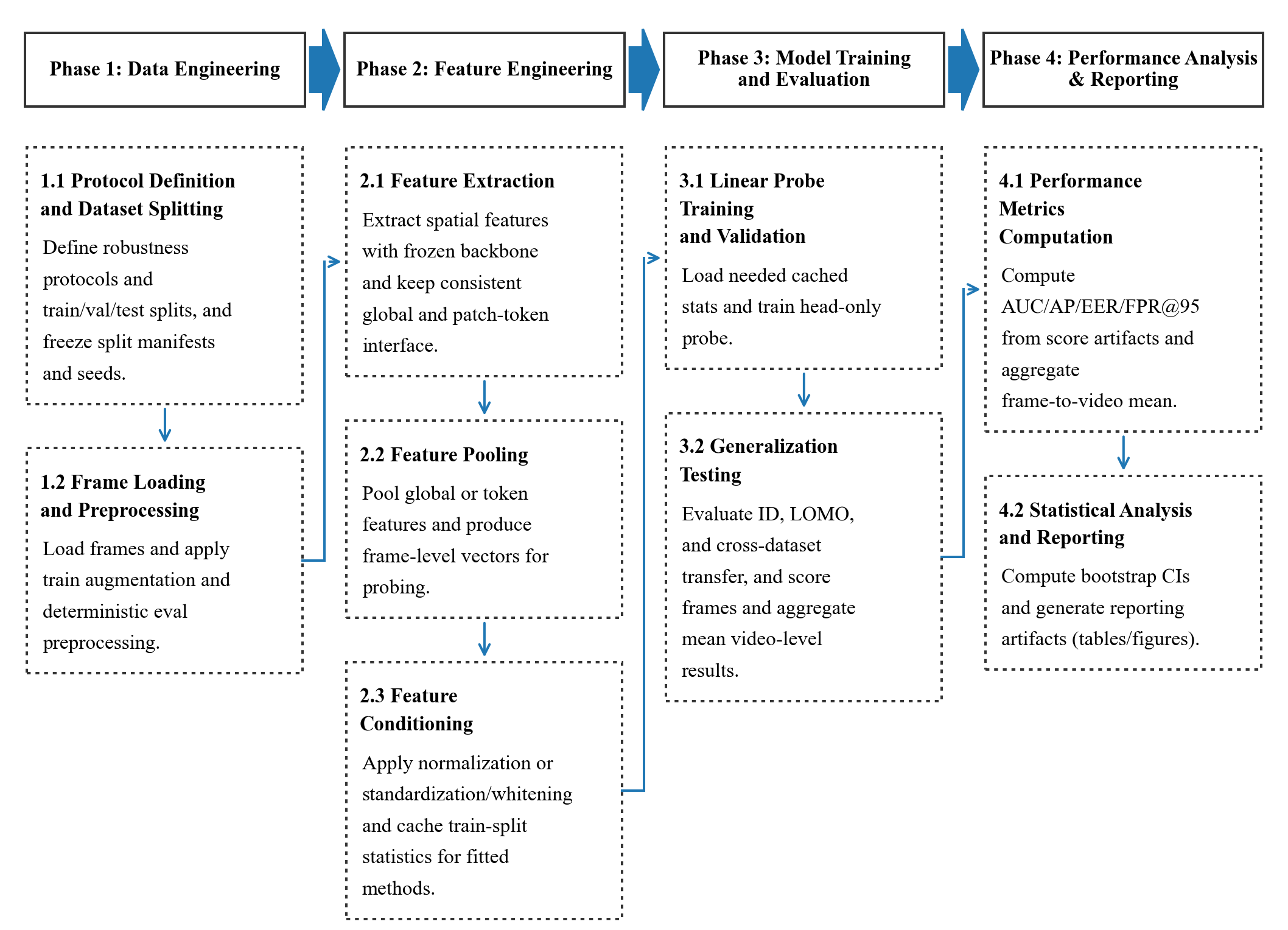}
\caption{Controlled probing framework for assessing robustness of frozen Vision Foundation Models under distribution shift.}
\label{fig:main_pipeline}
\end{figure*}

This paper isolates one variable: post-pool feature conditioning before a linear probe. The controlled setup keeps backbone, data protocol, and training recipe fixed. Figure~\ref{fig:main_pipeline} shows the pipeline.

The core research questions are: Does inherited conditioning help or hurt robustness under shift? Can conditioning alter ranking between ID and OOD protocols? We answer these using a controlled probing framework on FaceForensics++ c23 with transfer to Celeb-DF v2 and DFD.

Our main findings are: (1) conditioning materially changes robustness under shift, (2) ID-optimal ranking does not reliably transfer to OOD, and (3) robust choice can be protocol-dependent at higher backbone capacity. This supports treating feature conditioning as a primary design variable.

\section{Related Work}
The DINO line introduced self-distillation without labels in a teacher--student
framework and showed that self-supervised ViT features can encode strong
semantic structure \cite{caron2021dino}. DINOv2 scaled this recipe through
curated large-scale data and stronger frozen-feature transfer
\cite{oquab2024dinov2}. DINOv3 further scales to a ViT-7B teacher trained on
LVD-1689M and adds Gram anchoring to stabilize dense feature quality over long
training schedules \cite{simeoni2025dinov3}. It also provides heterogeneous
distillation into ConvNeXt students (Tiny/Small/Base/Large), enabling lower
inference-cost CNN backbones while preserving teacher-aligned representations.
CLIP and DINO families have made frozen-feature probing practical at scale
\cite{radford2021clip,caron2021dino,oquab2024dinov2,simeoni2025dinov3}, and
deepfake-focused CLIP studies report strong transfer potential
\cite{ojha2023ucf,yermakov2025unlocking}. However, these studies typically use default
frozen embeddings and do not isolate post-extraction conditioning as an
independent variable. To our knowledge, no prior work has applied DINOv3
ConvNeXt backbones to forged media detection.

In forged-media benchmarks, FaceForensics++ and DeepfakeBench show that high
in-distribution performance can coexist with weak cross-manipulation robustness
\cite{rossler2019faceforensics,yan2023deepfakebench}. Methods based on artifact
localization, reconstruction, and augmentation improve specific regimes but do
not remove shift sensitivity
\cite{li2020facexray,cao2022recce,shiohara2022sbi,yan2024lsda}, while Celeb-DF
further highlights realism-driven generalization gaps \cite{li2020celebdf}.
LayerNorm and BatchNorm are established normalization modules
\cite{ba2016layernorm,ioffe2015batchnorm}, and PCA whitening is a classic
retrieval transform \cite{jegou2012pca}. The methodological gap addressed here
is therefore isolation of post-extraction feature conditioning under fixed
backbone, data protocol, and optimizer settings, rather than changing
architecture or loss design.

\section{Methodology}

\subsection{Backbone Architecture: DINOv3 ConvNeXt}
The frozen backbones in this study are DINOv3 ConvNeXt models. Relative to
DINOv2, DINOv3 scales self-distillation to larger data and model regimes,
including pretraining on LVD-1689M (about 1.7B images) with a ViT-7B teacher
under a no-label objective~\cite{oquab2024dinov2,simeoni2025dinov3}. The
ConvNeXt student family (ConvNeXt-Tiny, ConvNeXt-Small, ConvNeXt-Base,
ConvNeXt-Large; 29M--198M parameters) is
obtained through heterogeneous distillation from that teacher. DINOv3 also
introduces Gram anchoring to stabilize dense feature learning over long training
schedules, improving patch-level representation consistency.
For compact table and figure captions, we abbreviate ConvNeXt-Tiny and
ConvNeXt-Base as CNX-Tiny and CNX-Base.
Architecturally, the ConvNeXt students retain hierarchical convolutional stages
with LayerNorm and inverted bottleneck-style blocks, while the ViT teacher uses
global token interactions via self-attention. Distillation aligns these
cross-architecture representations at lower inference cost. DINOv3 ConvNeXt
models are distilled from a ViT-7B teacher into efficient CNN backbones
explicitly designed for on-device deployment~\cite{simeoni2025dinov3}, making
them natural candidates for edge AI sensing pipelines. In our frozen-backbone
pipeline, only a linear probe is trained on 768-d (ConvNeXt-Tiny) or 1024-d
(ConvNeXt-Base)
descriptors. To our knowledge, no prior work has applied DINOv3 ConvNeXt
backbones to forged media detection; this work establishes conditioning
baselines, and absolute AUC comparisons to fine-tuned ViT-based detectors are
outside scope.

\subsection{Probing Framework}
An overview of the controlled probing pipeline is shown in
Fig.~\ref{fig:main_pipeline}. To ensure reproducible and fair evaluation of
frozen VFMs for forged media detection, we use a controlled framework spanning
data engineering, feature engineering, model training and evaluation, and
performance analysis.

The pipeline starts with protocol definition and split control. We define ID,
LOMO, and cross-dataset protocols, fix train/val/test splits, and freeze split
manifests and seeds with integrity checks. Frames are loaded by video decode or
frame manifests, with stochastic augmentation during training and deterministic
preprocessing for validation/test. For FaceForensics++ c23 (ID), fixed splits
are train 107,967 real / 91,891 fake (199,858 total), validation 20,949 /
17,877 (38,826 total), and test 21,096 / 17,909 (39,005 total). For
cross-dataset test-only evaluation, Celeb-DF v2 has 178 real and 178 fake
videos (356 total), and DFD has 363 real and 363 fake videos (726 total).

Feature engineering extracts spatial features from the frozen backbone, pools
global or token features into frame-level vectors, and applies conditioning
before probing. Conditioning includes normalization and
standardization/whitening; for fitted transforms, train-split statistics are
pre-computed once and cached for validation/test reuse. LN-Affine is the
default ConvNeXt head output with built-in normalization and serves as the
natural baseline. No intermediate feature bank is persisted; features are
computed on-the-fly and passed directly to conditioning and probing. For the
global single-head LN-Affine variant, pooling and conditioning are
operationally coupled via \texttt{\_extract\_global(norm=True)}.

A head-only linear probe is trained on frozen, conditioned features, with
checkpoint selection by best validation AUC across epochs. Generalization is
evaluated under ID, OOD cross-manipulation (LOMO folds), and OOD cross-dataset
transfer without retraining, with frame-level scores aggregated by mean to
video-level scores.

Finally, AUC, AP, EER, and FPR@95 are computed from saved score artifacts, and
bootstrap confidence intervals are derived from saved scores without retraining
to produce reporting artifacts. Across the full pipeline, backbone,
optimization recipe, and protocol are fixed; only conditioning is varied.

\section{Results}
Table~\ref{tab:03} reports in-distribution performance on FaceForensics++ c23.
LN is ID-optimal with a test AUC of 0.798, followed by LN-Affine at 0.784; the
full spread across conditioning choices is 9.3 AUC points. EER and FPR@95
remain high for all variants, indicating useful ranking signal but non-trivial
class overlap at strict operating points.

\begin{table}[!t]
\caption{In-distribution (ID) results. Best run by validation AUC.}
\label{tab:03}
\centering
\footnotesize
\setlength{\tabcolsep}{2pt}
\makebox[\columnwidth][c]{\textbf{(a) Validation metrics for CNX-Tiny}}
\vspace{2pt}
\begin{tabular}{lcccc}
\toprule
Condition & AUC & AP & EER & FPR@95 \\
\midrule
LN & 0.790 & 0.752 & 0.264 & 0.579 \\
LN-Affine & 0.776 & 0.742 & 0.279 & 0.679 \\
PCA-Whiten & 0.768 & 0.721 & 0.307 & 0.600 \\
Feature-Std & 0.758 & 0.719 & 0.307 & 0.707 \\
L2 & 0.705 & 0.689 & 0.343 & 0.843 \\
\bottomrule
\end{tabular}

\vspace{4pt}
\makebox[\columnwidth][c]{\textbf{(b) Test metrics for CNX-Tiny}}
\vspace{2pt}
\begin{tabular}{lcccc}
\toprule
Condition & AUC (95\% CI) & AP (95\% CI) & EER & FPR@95 \\
\midrule
LN & 0.798 (0.737-0.847) & 0.785 (0.723-0.837) & 0.257 & 0.650 \\
LN-Affine & 0.784 (0.726-0.835) & 0.752 (0.683-0.815) & 0.279 & 0.636 \\
PCA-Whiten & 0.777 (0.718-0.831) & 0.754 (0.685-0.819) & 0.286 & 0.757 \\
Feature-Std & 0.760 (0.695-0.813) & 0.734 (0.665-0.794) & 0.314 & 0.693 \\
L2 & 0.705 (0.649-0.762) & 0.679 (0.612-0.746) & 0.336 & 0.764 \\
\bottomrule
\end{tabular}
\end{table}

\begin{figure}[!t]
\centering
\includegraphics[width=\columnwidth]{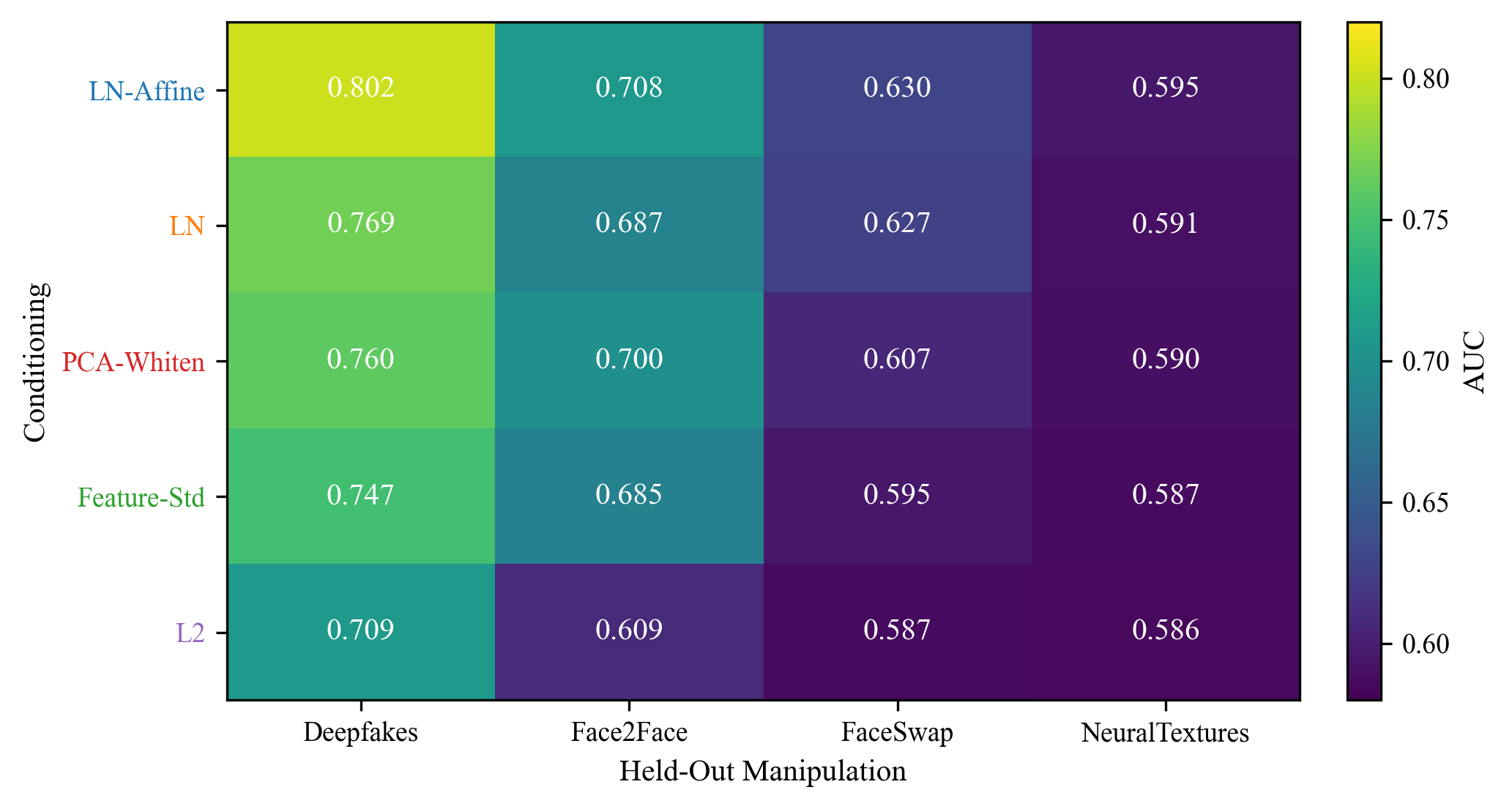}
\caption{LOMO AUC heatmap for CNX-Tiny across conditioning variants and held-out manipulations.}
\label{fig:main_lomo_heatmap}
\end{figure}

\begin{figure}[!t]
\centering
\includegraphics[width=\columnwidth]{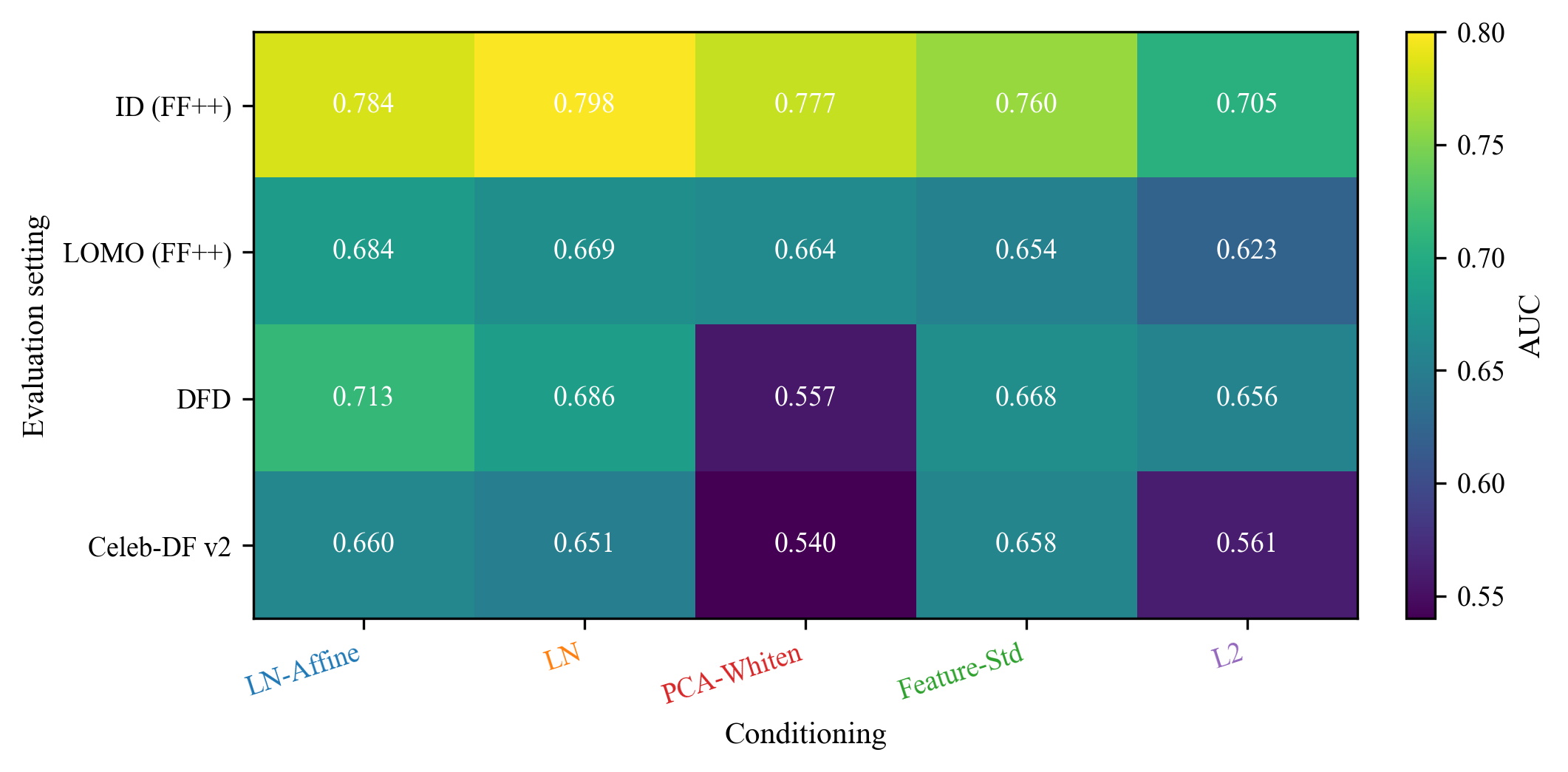}
\caption{Protocol-level AUC heatmap for CNX-Tiny (ID, LOMO, DFD, Celeb-DF v2).}
\label{fig:main_protocol_heatmap}
\end{figure}

\begin{figure}[!t]
\centering
\includegraphics[width=\columnwidth]{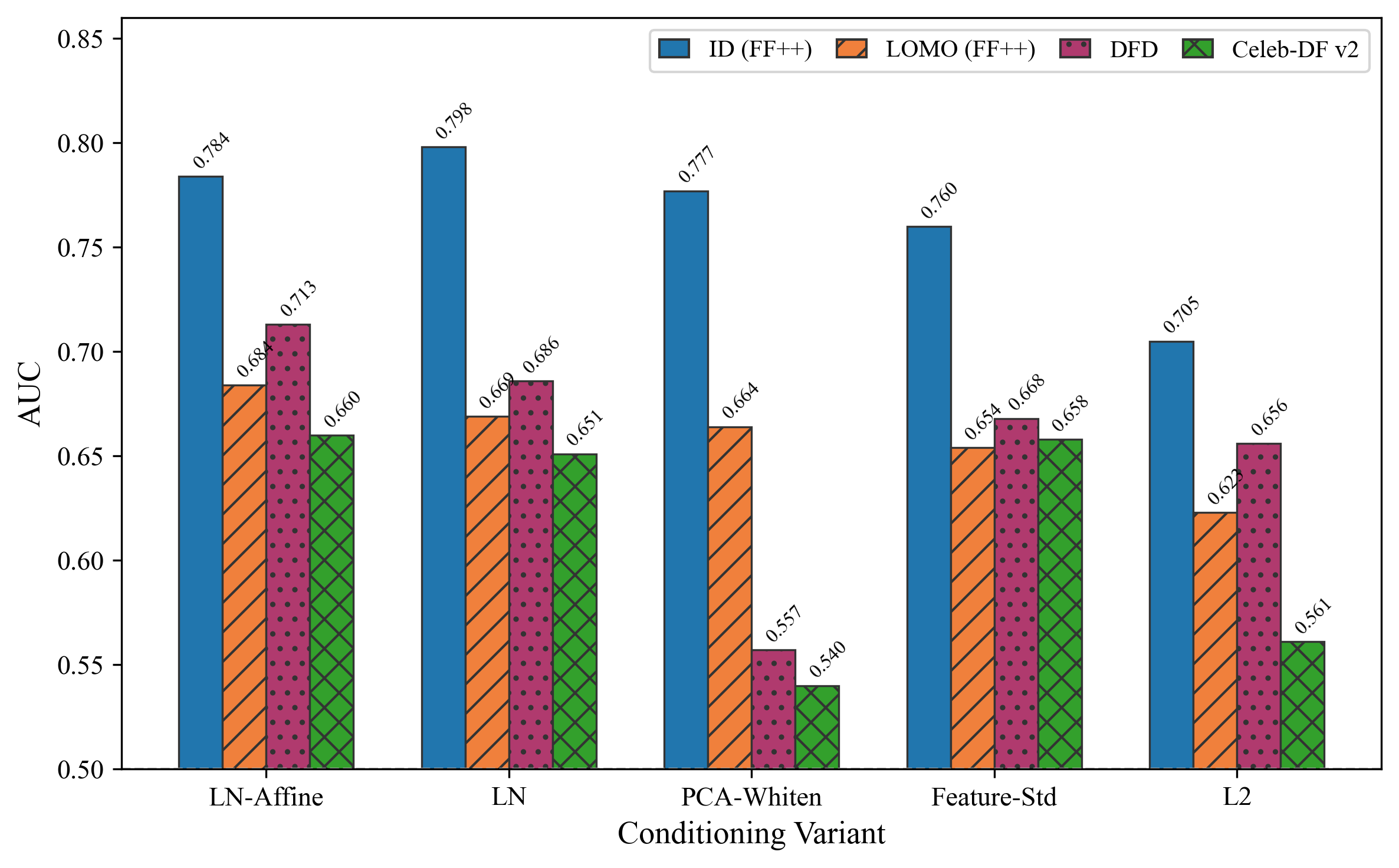}
\caption{Protocol-wise AUC summary for CNX-Tiny.}
\label{fig:main_tiny_summary}
\end{figure}

Under distribution shift, however, the ranking changes. The fold-averaged LOMO
summary in Table~\ref{tab:04} shows LN-Affine as the best mean OOD choice
(0.684), above LN (0.669), reversing the ID ordering. Per-fold detail in
Table~\ref{tab:05} shows the advantage is strongest on Deepfakes and Face2Face,
with convergence on the harder FaceSwap and NeuralTextures folds.
Fig.~\ref{fig:main_lomo_heatmap} visualizes this LOMO pattern for
ConvNeXt-Tiny.
Cross-dataset transfer in Table~\ref{tab:08} confirms the same pattern:
LN-Affine leads on both Celeb-DF v2 (0.660) and DFD (0.713), while
PCA-Whiten, competitive within-dataset, drops sharply under true dataset shift
(0.540 on Celeb-DF v2), consistent with train-domain covariance mismatch in
fitted whitening. Table~\ref{tab:06} provides full metrics for these external
datasets. Figs.~\ref{fig:main_protocol_heatmap} and \ref{fig:main_tiny_summary}
summarize the protocol-level AUC landscape for ConvNeXt-Tiny across all
evaluation settings.

\begin{table}[t]
\caption{FOLD-AVERAGED LOMO SUMMARY for CNX-Tiny}
\label{tab:04}
\centering
\footnotesize
\begin{tabular}{lcc}
\toprule
Condition & LOMO Mean AUC (95\% CI) & LOMO Std AUC \\
\midrule
LN-Affine & 0.684 (0.654-0.715) & 0.092 \\
LN & 0.669 (0.638-0.698) & 0.078 \\
PCA-Whiten & 0.664 (0.633-0.694) & 0.080 \\
Feature-Std & 0.654 (0.624-0.686) & 0.077 \\
L2 & 0.623 (0.590-0.655) & 0.059 \\
\bottomrule
\end{tabular}
\end{table}

\begin{table}[!t]
\caption{Per-Fold LOMO AUC Results for CNX-Tiny}
\label{tab:05}
\centering
\scriptsize
\setlength{\tabcolsep}{2.5pt}
\renewcommand{\arraystretch}{0.92}
\resizebox{\columnwidth}{!}{%
\begin{tabular}{llccc}
\toprule
Condition & Held-out & Val AUC & Test AUC (95\% CI) & Test AP (95\% CI) \\
\midrule
LN-Affine & DF & 0.782 & 0.802 (0.748-0.852) & 0.822 (0.774-0.865) \\
LN-Affine & F2F & 0.801 & 0.708 (0.647-0.772) & 0.699 (0.636-0.769) \\
LN-Affine & FS & 0.836 & 0.630 (0.562-0.694) & 0.644 (0.575-0.711) \\
LN-Affine & NT & 0.849 & 0.595 (0.529-0.665) & 0.578 (0.521-0.654) \\
LN & DF & 0.741 & 0.769 (0.717-0.820) & 0.773 (0.717-0.826) \\
LN & F2F & 0.789 & 0.687 (0.627-0.749) & 0.674 (0.610-0.742) \\
LN & FS & 0.829 & 0.627 (0.564-0.692) & 0.632 (0.568-0.693) \\
LN & NT & 0.812 & 0.591 (0.524-0.655) & 0.572 (0.513-0.645) \\
PCA-Whiten & DF & 0.735 & 0.760 (0.699-0.812) & 0.776 (0.719-0.825) \\
PCA-Whiten & F2F & 0.773 & 0.700 (0.637-0.757) & 0.705 (0.641-0.766) \\
PCA-Whiten & FS & 0.799 & 0.607 (0.544-0.671) & 0.582 (0.521-0.652) \\
PCA-Whiten & NT & 0.809 & 0.590 (0.528-0.652) & 0.570 (0.516-0.642) \\
Feature-Std & DF & 0.743 & 0.747 (0.693-0.799) & 0.769 (0.709-0.822) \\
Feature-Std & F2F & 0.794 & 0.685 (0.623-0.746) & 0.673 (0.613-0.740) \\
Feature-Std & FS & 0.820 & 0.595 (0.522-0.661) & 0.595 (0.528-0.665) \\
Feature-Std & NT & 0.838 & 0.587 (0.521-0.656) & 0.574 (0.515-0.649) \\
L2 & DF & 0.655 & 0.709 (0.645-0.770) & 0.713 (0.650-0.775) \\
L2 & F2F & 0.708 & 0.609 (0.546-0.681) & 0.600 (0.534-0.675) \\
L2 & FS & 0.724 & 0.587 (0.519-0.653) & 0.589 (0.526-0.658) \\
L2 & NT & 0.720 & 0.586 (0.520-0.655) & 0.574 (0.516-0.646) \\
\bottomrule
\end{tabular}
}
\end{table}

\begin{table}[t]
\caption{COMBINED CROSS-DATASET SUMMARY for CNX-Tiny}
\label{tab:08}
\centering
\footnotesize
\begin{tabular}{lccc}
\toprule
Condition & Celeb-DF v2 AUC & DFD AUC & Mean XD AUC \\
\midrule
LN-Affine & 0.660 & 0.713 & 0.687 \\
LN & 0.651 & 0.686 & 0.669 \\
Feature-Std & 0.658 & 0.668 & 0.663 \\
L2 & 0.561 & 0.656 & 0.609 \\
PCA-Whiten & 0.540 & 0.557 & 0.549 \\
\bottomrule
\end{tabular}
\end{table}

\begin{table}[!t]
\caption{Cross-dataset results on Celeb-DF v2 and DFD.}
\label{tab:06}
\centering
\footnotesize
\setlength{\tabcolsep}{2pt}
\textbf{(a) Celeb-DF v2 for CNX-Tiny}
\vspace{2pt}
\begin{tabular}{lcccc}
\toprule
Condition & AUC (95\% CI) & AP (95\% CI) & EER & FPR@95 \\
\midrule
LN-Affine & 0.660 (0.607-0.715) & 0.610 (0.555-0.674) & 0.371 & 0.899 \\
LN & 0.651 (0.592-0.707) & 0.619 (0.561-0.681) & 0.399 & 0.854 \\
Feature-Std & 0.658 (0.597-0.713) & 0.611 (0.558-0.676) & 0.354 & 0.871 \\
L2 & 0.561 (0.505-0.626) & 0.522 (0.479-0.583) & 0.461 & 0.888 \\
PCA-Whiten & 0.540 (0.480-0.599) & 0.559 (0.504-0.622) & 0.483 & 0.989 \\
\bottomrule
\end{tabular}

\vspace{4pt}
\textbf{(b) DFD for CNX-Tiny}
\vspace{2pt}
\begin{tabular}{lcccc}
\toprule
Condition & AUC (95\% CI) & AP (95\% CI) & EER & FPR@95 \\
\midrule
LN-Affine & 0.713 (0.674-0.748) & 0.680 (0.635-0.722) & 0.353 & 0.738 \\
LN & 0.686 (0.649-0.723) & 0.664 (0.624-0.711) & 0.353 & 0.893 \\
Feature-Std & 0.668 (0.630-0.710) & 0.640 (0.598-0.688) & 0.361 & 0.895 \\
L2 & 0.656 (0.615-0.693) & 0.650 (0.609-0.693) & 0.402 & 0.796 \\
PCA-Whiten & 0.557 (0.518-0.600) & 0.564 (0.526-0.608) & 0.455 & 0.912 \\
\bottomrule
\end{tabular}
\end{table}

\begin{figure}[!t]
\centering
\includegraphics[width=\columnwidth]{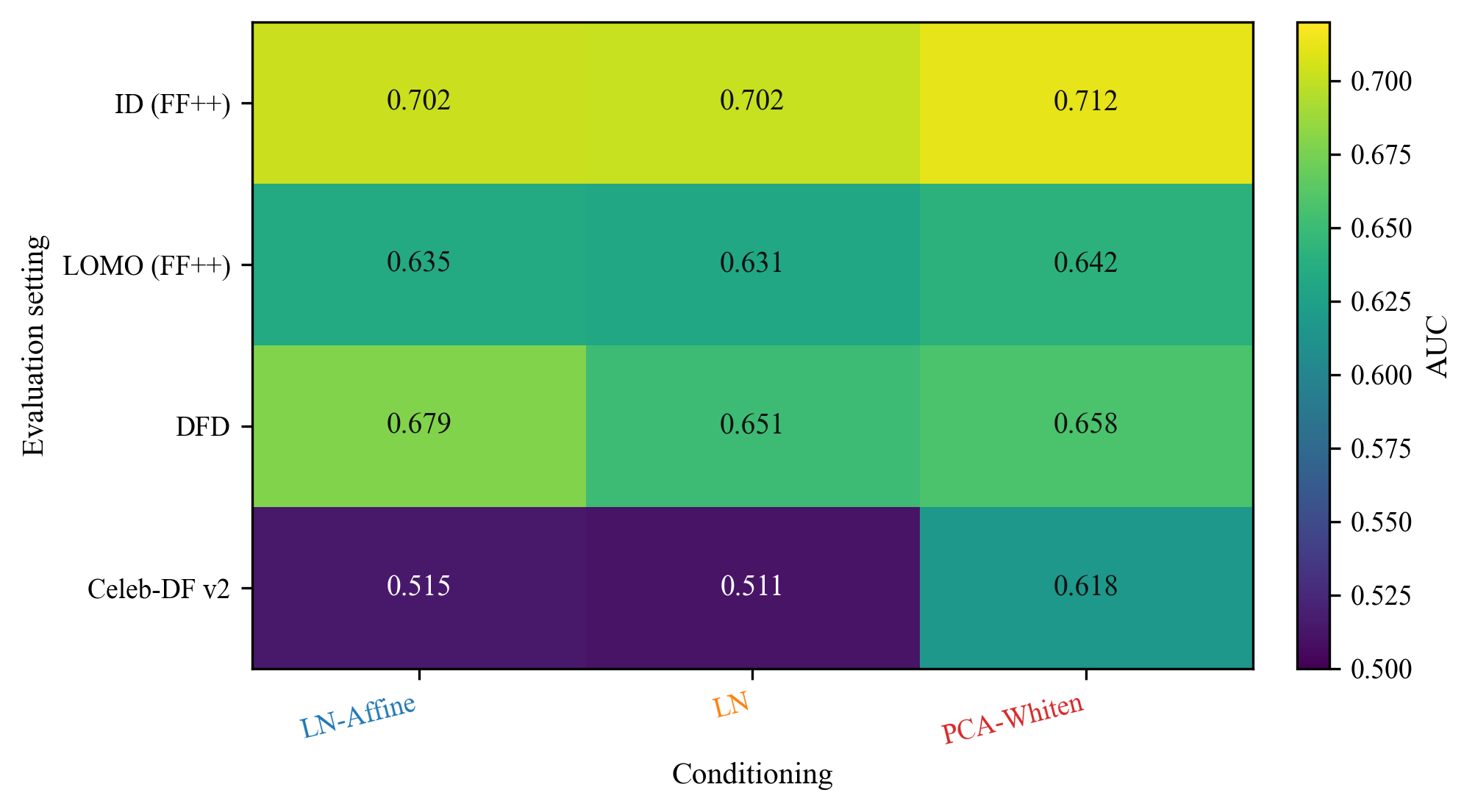}
\caption{Protocol-level AUC heatmap for CNX-Base.}
\label{fig:main_base_heatmap}
\end{figure}

The ConvNeXt-Base replication (Table~\ref{tab:09}) tests whether these patterns
hold at higher capacity. We replicate the top-3 conditioning variants ranked by
ConvNeXt-Tiny LOMO mean AUC; this compute-bounded scope is not selective
reporting because the full five-variant sweep is reported for ConvNeXt-Tiny
(Tables~\ref{tab:03}--\ref{tab:06}). ConvNeXt-Base does not preserve a single
OOD winner across protocols: PCA-Whiten leads on ID (0.712), LOMO (0.642), and
Celeb-DF v2 (0.618), while LN-Affine leads on DFD (0.679).
Fig.~\ref{fig:main_base_heatmap} shows the Base protocol-level heatmap. This
protocol-dependent outcome at higher capacity reinforces the core finding:
robust conditioning choice cannot be derived from ID performance alone.
Table~\ref{tab:10} compares protocol winners across both backbones, providing
the synthesis view that the ID-optimal condition is never the universal OOD
winner.

\begin{table}[t]
\caption{CNX-Base Replication AUC Results}
\label{tab:09}
\centering
\footnotesize
\begin{tabular}{lcccc}
\toprule
Conditioning & ID & LOMO & Celeb-DF v2 & DFD \\
\midrule
LN-Affine & 0.702 & 0.635 & 0.515 & 0.679 \\
LN & 0.702 & 0.631 & 0.511 & 0.651 \\
PCA-Whiten & 0.712 & 0.642 & 0.618 & 0.658 \\
\bottomrule
\end{tabular}
\end{table}

\begin{table}[!ht]
\caption{Protocol winner by backbone (AUC) (Best Conditioning Variant per Protocol).}
\label{tab:10}
\centering
\footnotesize
\begin{tabular}{lll}
\toprule
Protocol & CNX-Tiny winner & CNX-Base winner \\
\midrule
ID & LN (0.798) & PCA-Whiten (0.712) \\
LOMO & LN-Affine (0.684) & PCA-Whiten (0.642) \\
Celeb-DF v2 & LN-Affine (0.660) & PCA-Whiten (0.618) \\
DFD & LN-Affine (0.713) & LN-Affine (0.679) \\
\bottomrule
\end{tabular}
\end{table}

\begin{figure}[!t]
\centering
\includegraphics[width=0.71\columnwidth]{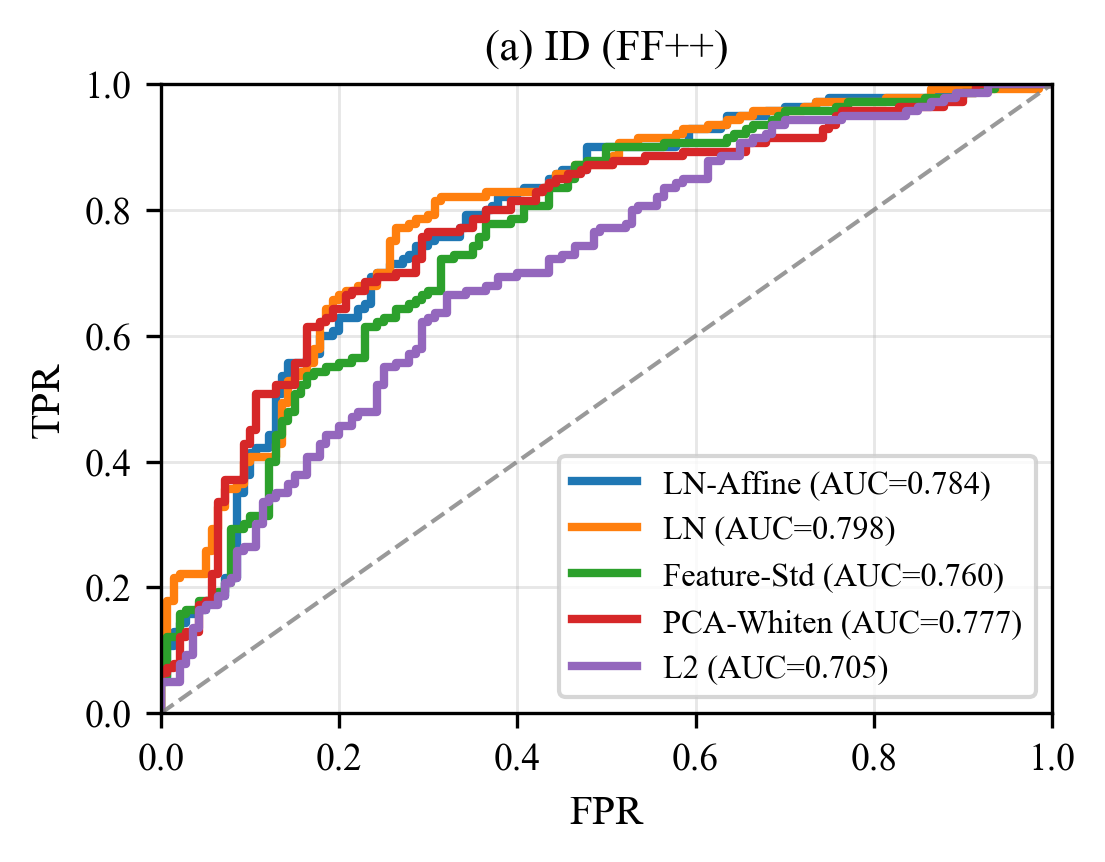}\\[-1pt]
\includegraphics[width=0.71\columnwidth]{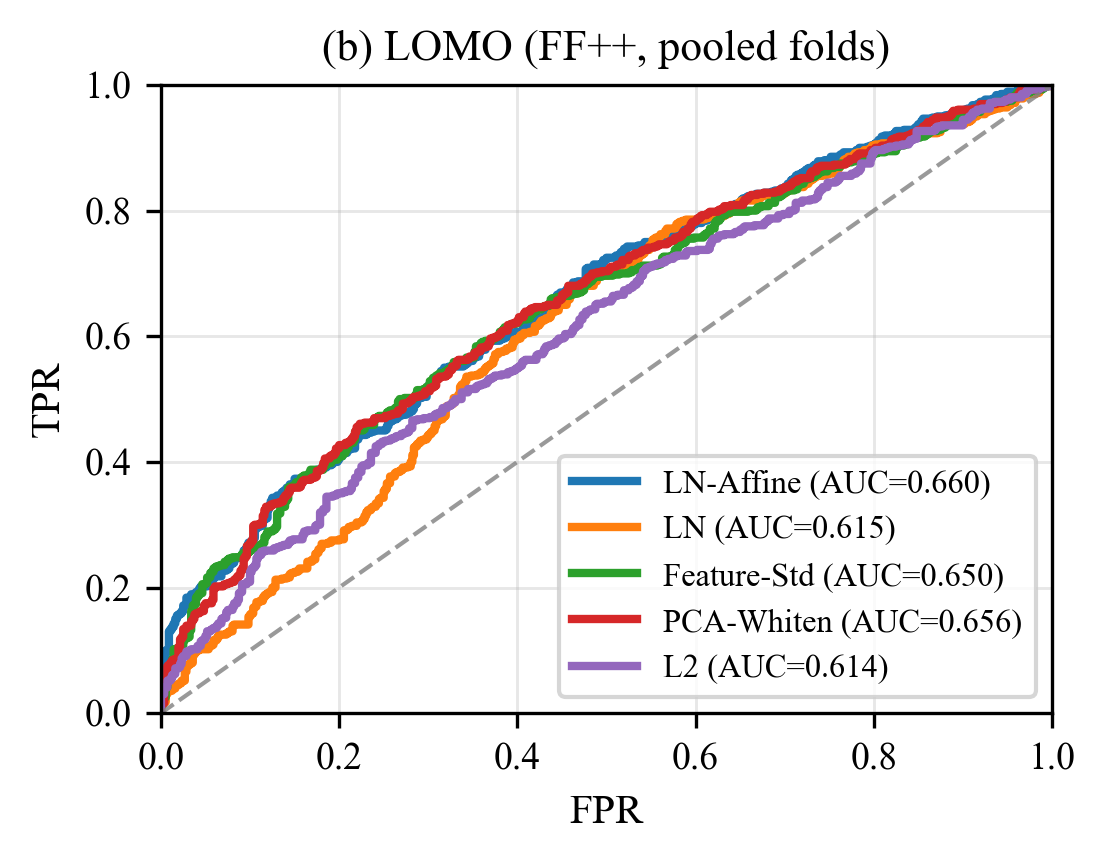}
\caption{ROC curves for CNX-Tiny across ID and LOMO protocols.}
\label{fig:main_roc}
\end{figure}

\begin{figure}[!t]
\centering
\includegraphics[width=0.71\columnwidth]{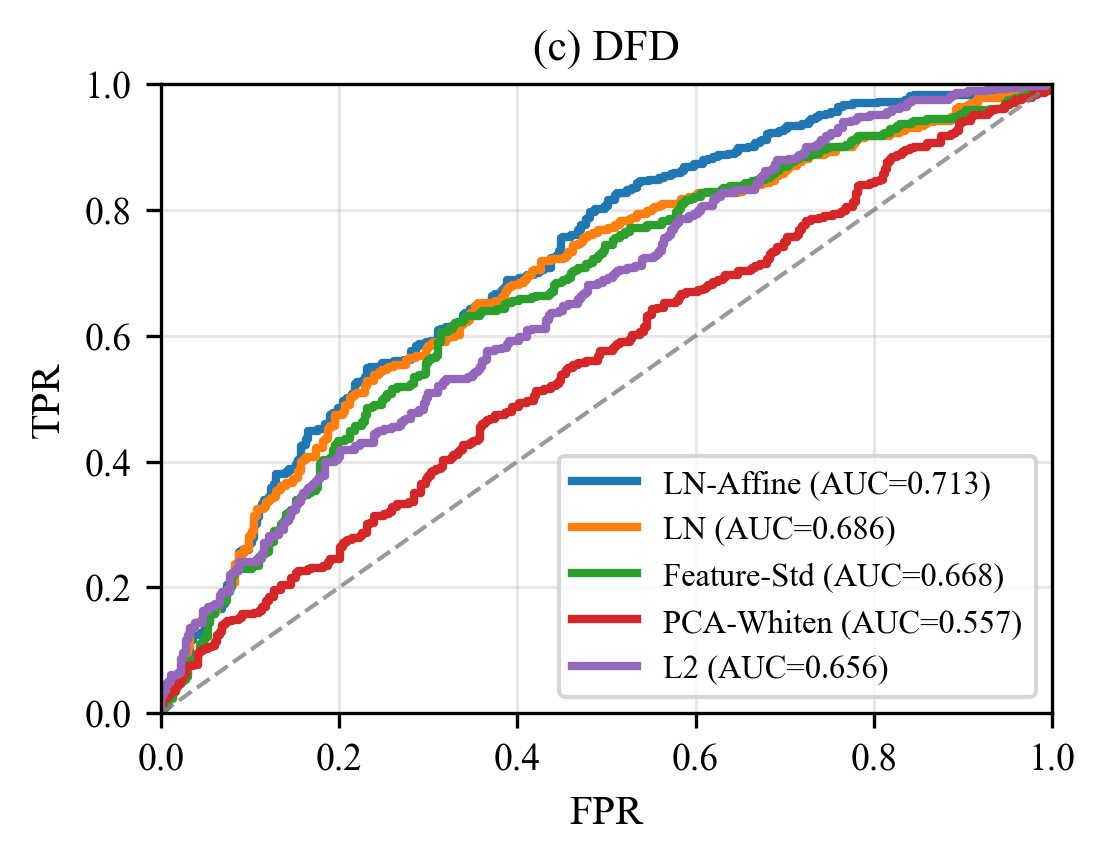}\\[-1pt]
\includegraphics[width=0.71\columnwidth]{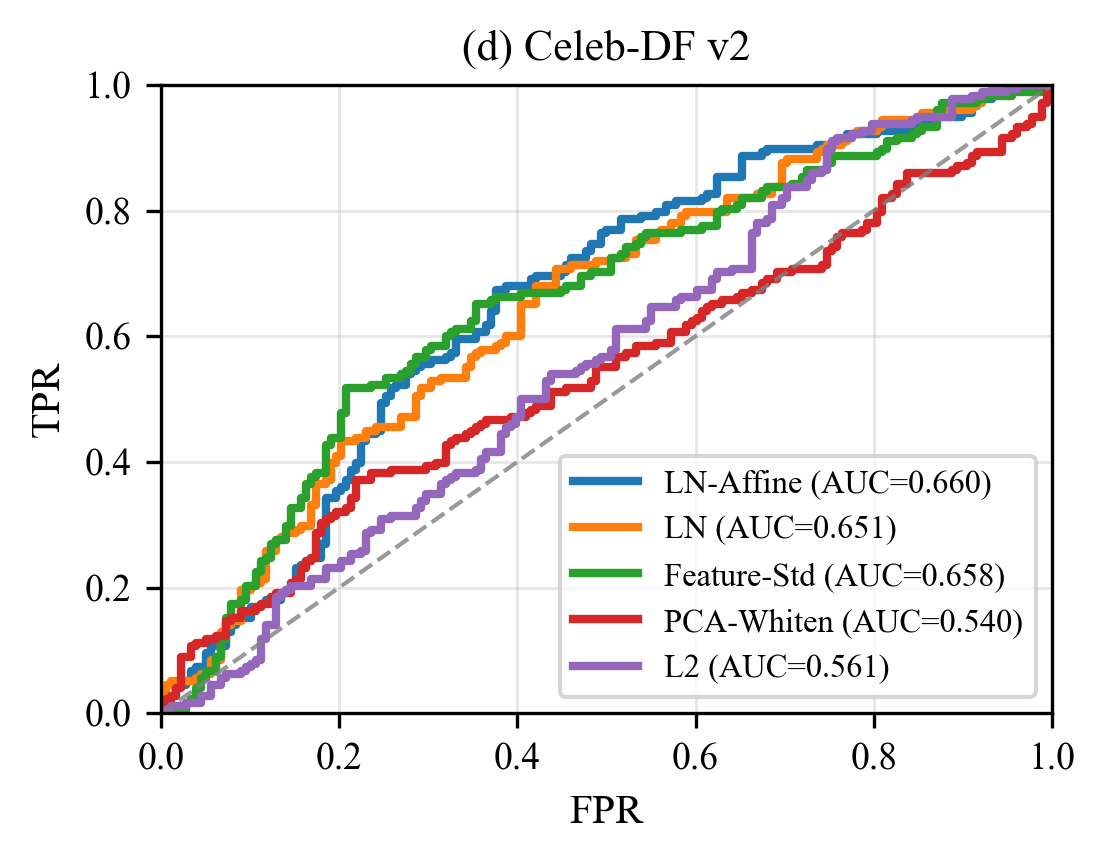}
\caption{ROC curves for CNX-Tiny across cross-dataset protocols (DFD and Celeb-DF v2).}
\label{fig:main_roc_ood}
\end{figure}

\FloatBarrier

\begin{figure}[!t]
\centering
\includegraphics[width=0.71\columnwidth]{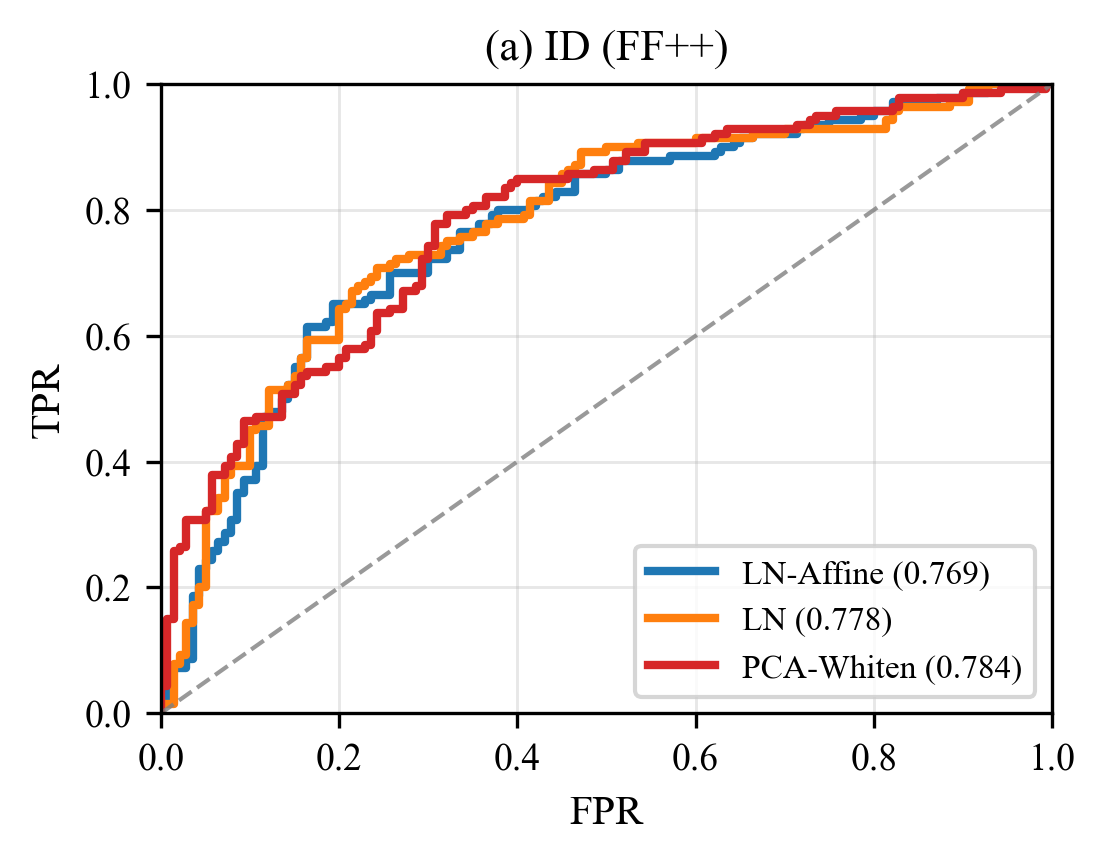}\\[-1pt]
\includegraphics[width=0.71\columnwidth]{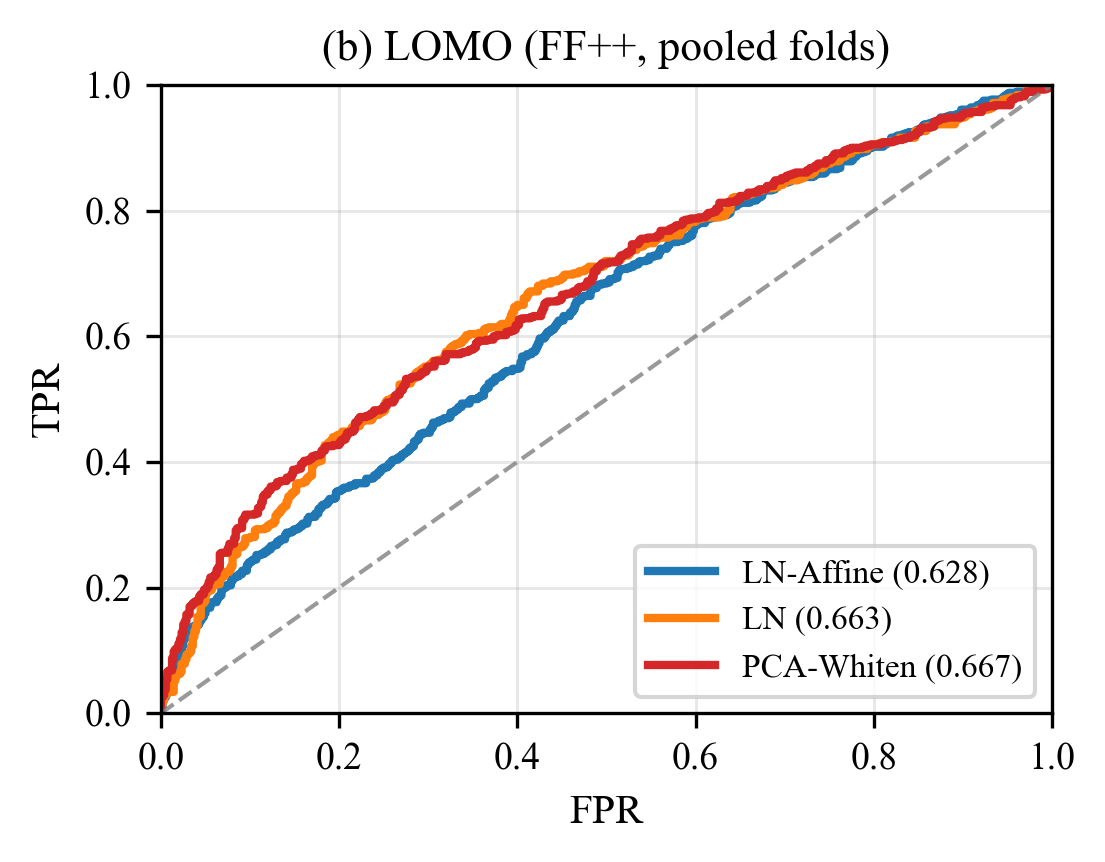}
\caption{ROC curves for CNX-Base across ID and LOMO protocols.}
\label{fig:main_base_roc}
\end{figure}

\begin{figure}[!t]
\centering
\includegraphics[width=0.71\columnwidth]{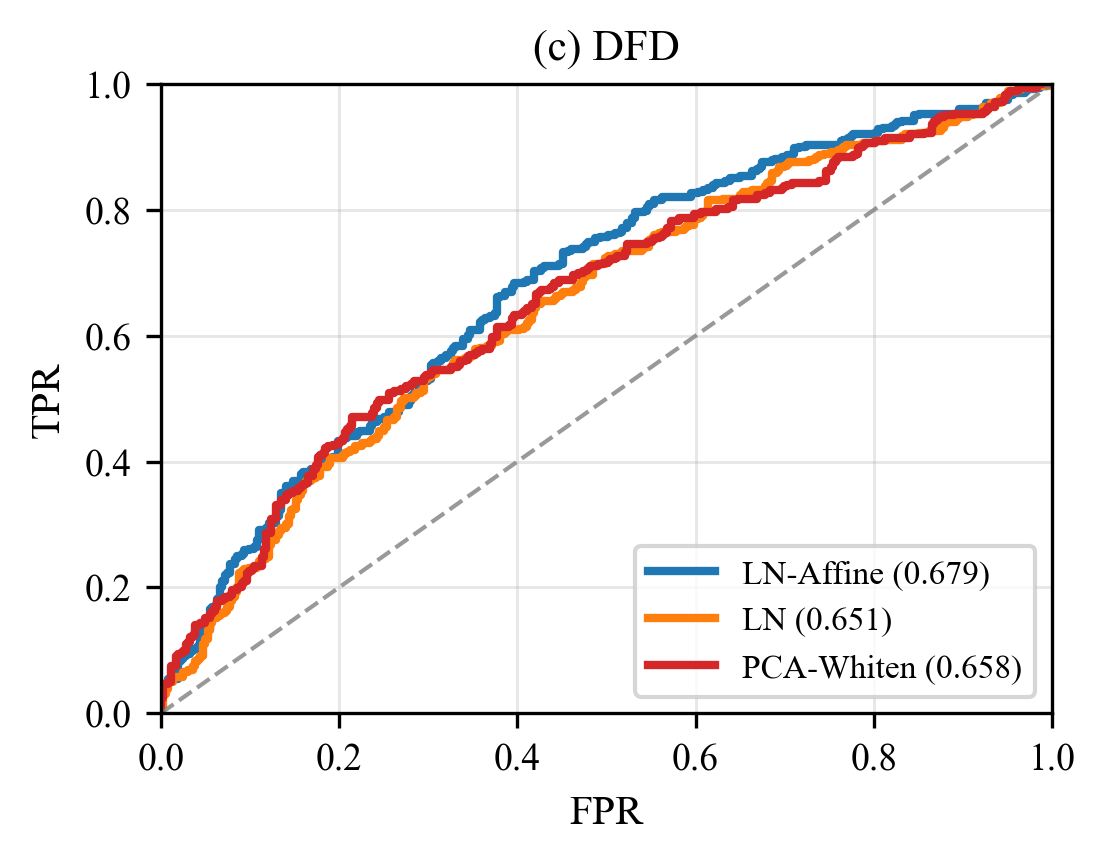}\\[-1pt]
\includegraphics[width=0.71\columnwidth]{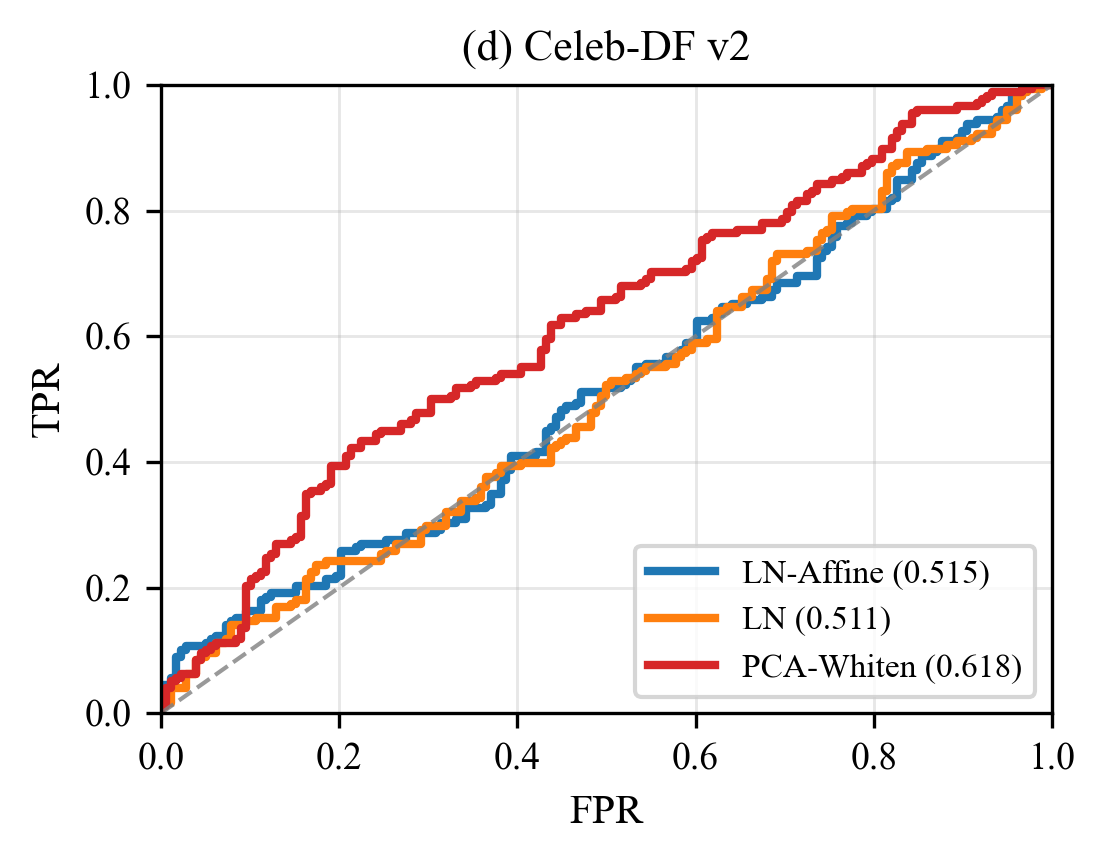}
\caption{ROC curves for CNX-Base across cross-dataset protocols (DFD and Celeb-DF v2).}
\label{fig:main_base_roc_ood}
\end{figure}

Two patterns are consistent across all analyses. First, ID-vs-OOD misalignment:
the ID-optimal selection does not reliably transfer under manipulation or
dataset shift. Second, conditioning has a material effect: a
zero-backbone-retraining interface choice changes robustness without changing
architecture or data protocol. This implies conditioning should be tuned against
target shift, not selected by ID accuracy alone.

Figs.~\ref{fig:main_roc}, \ref{fig:main_roc_ood}, \ref{fig:main_base_roc}, and
\ref{fig:main_base_roc_ood} present ROC curves for ConvNeXt-Tiny and
ConvNeXt-Base across all evaluation protocols, and they show clear threshold
dependence. In LOMO, LN-Affine and LN cross by operating point;
Fig.~\ref{fig:main_roc}(b) reports pooled out-of-fold AUCs, whereas
Table~\ref{tab:04} reports mean per-fold video-level AUCs, so the values are
related but not identical. In cross-dataset transfer, PCA-Whiten moves closer
to the diagonal on ConvNeXt-Tiny/Celeb-DF v2, consistent with train-domain
covariance mismatch; in ConvNeXt-Base/Celeb-DF v2 it recovers aggregate AUC but
can still underperform LN-Affine at low FPR. Thus AUC alone is insufficient for
deployment-oriented selection.

\section{Conclusion}
Feature conditioning at the frozen-model interface is not a harmless default. We present the first controlled probing study on DINOv3 ConvNeXt backbones -- heterogeneous distillates of a ViT-7B teacher designed for resource-constrained and edge AI deployment -- establishing performance baselines for this backbone family on face deepfake video detection where none previously existed. Under controlled probing, conditioning changes robustness materially and can reverse ID-vs-OOD ranking. Across protocols, DINOv3 ConvNeXt remains competitive under linear probing alone, showing that no task-specific backbone fine-tuning is required to obtain transferable forged-media detection features at edge-compatible cost. LN-Affine, the default ConvNeXt head output and natural baseline, proves competitive under external dataset transfer despite being suboptimal in-distribution, while PCA-Whiten -- competitive within-dataset -- degrades sharply under true dataset shift. Therefore, deployment-oriented model selection should use robustness-oriented validation (LOMO and external datasets), not ID accuracy alone.

The result is operationally relevant for frozen-feature pipelines targeting edge sensing applications: robustness failure can originate at the descriptor interface even when backbone and head are fixed. This creates a practical, zero-backbone-retraining lever for improving transfer behavior without retraining the backbone -- directly applicable to vehicular, industrial, and on-device authenticity verification systems where DINOv3 ConvNeXt's inference efficiency makes it a natural fit.

Future work includes broader multi-seed and multi-backbone sweeps. ConvNeXt-Base replication covers the top-3 variants by ConvNeXt-Tiny LOMO ranking; Feature-Std and L2 are excluded from Base runs as lower-ranked conditions under the reported Tiny protocols.

{\small
\bibliographystyle{IEEEtran}
\bibliography{refs/references_iwcmc}
}

\end{document}